\documentclass{article}

\usepackage{arxiv}

\usepackage[utf8]{inputenc} 
\usepackage[T1]{fontenc}    
\usepackage{hyperref}       
\usepackage{url}            
\usepackage{booktabs}       
\usepackage{amsfonts}       
\usepackage{nicefrac}       
\usepackage{microtype}      
\usepackage{graphicx}
\usepackage[numbers]{natbib}
\usepackage{doi}

\usepackage{float}
\usepackage{hyperref} 
\usepackage{amsmath} 

\usepackage{algorithm} 
\usepackage{algpseudocode}

\usepackage{rotating} 

\usepackage{listings}

\usepackage{xcolor}
\definecolor{codegreen}{rgb}{0,0.6,0}
\definecolor{codegray}{rgb}{0.5,0.5,0.5}
\definecolor{codepurple}{rgb}{0.58,0,0.82}
\definecolor{backcolour}{rgb}{1,1,1}
\lstdefinestyle{mystyle}{
    backgroundcolor=\color{backcolour},   
    commentstyle=\color{codegreen},
    keywordstyle=\color{magenta},
    numberstyle=\tiny\color{codegray},
    stringstyle=\color{codepurple},
    basicstyle=\ttfamily\footnotesize,
    breakatwhitespace=false,         
    breaklines=true,                 
    captionpos=t, 
    keepspaces=true,                 
    numbers=none, 
    numbersep=5pt,                  
    showspaces=false,                
    showstringspaces=false,
    showtabs=false,                  
    tabsize=2
}
\makeatletter
\pretocmd\lst@makecaption{\noindent{\rule{\linewidth}{1pt}}}{}{}
\makeatother

\title{Encrypted Vector Similarity Computations Using Partially Homomorphic Encryption: Applications and Performance Analysis}

\author{
    \href{https://orcid.org/0000-0002-0345-0088}{\includegraphics[scale=0.06]{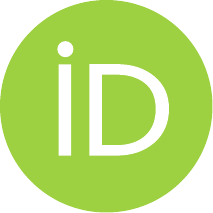}\hspace{1mm}Sefik Serengil} \\
	Solution Engineering\\
	Vorboss Limited\\
	London, UK \\
	\texttt{sefik.serengil@vorboss.com} \\
	\And
	\href{https://orcid.org/0000-0003-1250-5949}{\includegraphics[scale=0.06]{orcid.pdf}\hspace{1mm}Alper Ozpinar} \\
	Department of Business\\
	Ibn Haldun University\\
	Istanbul, Turkiye \\
	\texttt{alper@ozpinar.org} \\
}

\renewcommand{\shorttitle}{Encrypted Vector Similarity Computations Using PHE}

\hypersetup{
pdftitle={Encrypted Vector Similarity Computations Using Partially Homomorphic Encryption: Applications and Performance Analysis},
pdfsubject={cs.CR, cs.CV, cs.LG},
pdfauthor={Sefik Serengil, Alper Ozpinar},
pdfkeywords={Facial Recognition, Homomorphic Encryption, Similarity Search},
}

\begin{document}
\maketitle

\begin{abstract}
This paper investigates the application of partially homomorphic encryption (PHE) for encrypted vector similarity search, focusing on facial recognition but applicable to broader tasks such as reverse image search, recommendation engines, and large language models (LLMs). While fully homomorphic encryption (FHE) implementations exist, this work demonstrates that encrypted cosine similarity can be computed using PHE, providing a practical alternative due to its notable advantages. Since cosine similarity cannot be directly computed with PHE, we propose a method that normalizes both the source and target vectors in advance, enabling the use of dot product calculations as a proxy for cosine similarity. We also address challenges related to negative dimension values through min-max normalization. Our experiments, conducted on the Labeled Faces in the Wild (LFW) dataset with facial embeddings from DeepFace's FaceNet128d, FaceNet512d, and VGG-Face (4096d) models, adopt a two-tower architecture. In this setup, pre-encrypted vector embeddings are stored in one tower, while the other tower—represented by an edge device or mobile phone—captures facial images, computes their embeddings, and calculates the dot product between encrypted and plaintext vectors using additively homomorphic encryption. LightPHE is used to implement the additively homomorphic cryptosystem, evaluating encryption schemes such as Paillier, Damgård-Jurik, and Okamoto-Uchiyama. Other schemes like Benaloh, Naccache-Stern, Exponential-ElGamal, and Elliptic Curve ElGamal are excluded due to performance limitations or decryption complexity. The experiments run at 80-bit and 112-bit security levels, with 112-bit considered secure until 2030 according to NIST. We also compare PHE against FHE using the TenSEAL library, assessing metrics such as encryption time, decryption time, homomorphic operation time, cosine similarity loss upon decryption, key size, and ciphertext size. This study highlights that PHE, while less computationally intensive, faster, and producing smaller ciphertexts and keys, strikes a favorable balance for memory-constrained environments and practical use cases. The results underscore the potential of PHE for privacy-preserving encrypted vector similarity search across a variety of domains.
\end{abstract}

\keywords{Facial Recognition \and Homomorphic Encryption \and Similarity Search}

\section{Introduction}

Privacy-preserving machine learning has gained significant attention in recent years due to the increasing need to process sensitive data while ensuring confidentiality. One key challenge in this domain is performing vector similarity computations in an encrypted manner, which is crucial for applications such as facial recognition, reverse image search, recommendation systems, and large language model (LLM) retrieval mechanisms. Fully Homomorphic Encryption (FHE) \cite{fhe} has been proposed as a solution to enable computations on encrypted data \cite{cipherface}; however, FHE-based approaches suffer from high computational overhead, large ciphertext sizes, and impractical key management, making them difficult to deploy in real-world, resource-constrained environments.

In this work, we explore Partially Homomorphic Encryption (PHE) \cite{phe} as an alternative for encrypted vector similarity computation. Unlike FHE, which supports arbitrary computations on encrypted data, PHE provides a more lightweight approach by allowing a limited set of operations—addition and scalar multiplication or multiplication—while significantly reducing computational costs and storage requirements. Despite its limitations, we demonstrate that PHE can be effectively leveraged to perform encrypted vector similarity search by utilizing a normalization-based transformation that enables cosine similarity computation using only dot products.

To validate our approach, we focus on the facial recognition task using the Labeled Faces in the Wild (LFW) \cite{lfw} dataset. We extract vector embeddings from facial images using three widely used deep learning models from DeepFace \cite{deepface} \cite{deepface2}: FaceNet \cite{facenet} (128D and 512D) and VGG-Face \cite{vggface} (4096D). The embeddings of registered users are pre-encrypted using LightPHE \cite{lightphe}, a Python library supporting several additively homomorphic encryption schemes, including Paillier \cite{paillier}, Damgård-Jurik \cite{damgard}, and Okamoto-Uchiyama \cite{okamoto} cryptosystems. Other schemes like Benaloh \cite{benaloh}, Naccache-Stern \cite{naccache}, Exponential-ElGamal \cite{elgamal}, and Elliptic Curve ElGamal \cite{ecelgamal} are excluded due to performance limitations or decryption complexity. At query time, a facial image captured on an edge device is converted into a vector embedding and compared against the encrypted database using homomorphic dot product operations, which are feasible under additively homomorphic encryption.

Our contributions can be summarized as follows:

\begin{itemize}

\item We propose an efficient encrypted vector similarity computation method using Partially Homomorphic Encryption (PHE), demonstrating its feasibility as a lightweight alternative to FHE.

\item We introduce a normalization-based transformation that enables cosine similarity computation using only homomorphic dot products, overcoming the inherent limitations of PHE.

\item We conduct extensive experiments on facial recognition tasks over the LFW dataset, using three different embedding sizes to analyze performance across varying dimensionalities.

\item We conduct extensive experiments with different additively homomorphic encryption algorithms to compare them each other.

\item We benchmark the performance of PHE-based vector similarity against FHE-based methods (using TenSEAL \cite{tenseal}), comparing key metrics such as encryption time, decryption time, homomorphic operation time, cosine similarity loss upon decryption, key sizes, and average encrypted embedding and encrypted similarity sizes across 80-bit and 112-bit security levels.

\item Our results show that PHE significantly outperforms FHE in terms of efficiency, making it a viable solution for privacy-preserving similarity search on resource-constrained environments such as edge devices and mobile platforms. While FHE remains the most secure approach for fully arbitrary computations, PHE presents a compelling trade-off between security and efficiency, enabling practical encrypted similarity search for real-world applications.

\end{itemize}

The rest of this paper is organized as follows: Section 2 provides an overview of partially homomorphic encryption and discusses proofs of homomorphic addition and scalar multiplication in the Paillier, Damgård-Jurik, and Okamoto-Uchiyama algorithms. Section 3 introduces the cosine similarity formula used in this work, along with a modified version adapted for partial homomorphic encryption (PHE). Section 4 elaborates on the implementation of cosine similarity within the context of PHE. Section 5 describes the experimental setup in detail and presents the results of our evaluation, comparing PHE and fully homomorphic encryption (FHE) methods based on various performance metrics. Finally, Section 6 concludes the paper, highlighting key findings, limitations, and potential directions for future research.

\section{Partially Homomorphic Encryption}

Homomorphic encryption enables computations on encrypted data in such a way that, when decrypted, the result matches what would have been obtained if the operation were performed on the plaintext data. This feature is especially valuable in situations where data privacy needs to be preserved while still allowing computations to be carried out, such as in cloud-based systems.

In additive homomorphism, the encryption scheme allows for the addition of ciphertexts as shown in Equation \ref{eq:homomorphic_computation_add}.

\begin{equation}
\label{eq:homomorphic_computation_add}
    {D(E(A) + E(B)) = A + B}
\end{equation}

where \( E \) denotes encryption and \( D \) represents decryption.

Additively homomorphic schemes also supports scalar multiplication where you are able to multiply a ciphertext with a known plaintext as illustrated in Equation \ref{eq:homomorphic_computation_scalar}.

\begin{equation}
\label{eq:homomorphic_computation_scalar}
    {D(k \times E(A)) = k \times A}
\end{equation}

In multiplicative homomorphism, the encryption scheme enables the multiplication of ciphertexts. For instance, the RSA and ElGamal encryption schemes allow ciphertexts to be multiplied together, with the resulting ciphertext corresponding to the product of the plaintexts:

\begin{equation}
\label{eq:homomorphic_computation_multiply}
    {D(E(A) \cdot E(B)) = A \times B}
\end{equation}

An algorithm that exhibits either additive or multiplicative homomorphic properties is referred to as partially homomorphic encryption (PHE). In contrast, a fully homomorphic encryption (FHE) scheme supports both additive and multiplicative homomorphisms. This means it allows complex operations, such as addition and multiplication, to be performed on encrypted data, offering greater flexibility in computations while preserving the data's confidentiality.

\subsection{Paillier}

The Paillier cryptosystem was introduced by Pascal Paillier in 1999 \cite{paillier}. It relies on the computational difficulty of determining n-th residue classes. The general encryption process of the Paillier algorithm is presented in Equation \ref{eq:paillier_generic}, where m represents the plaintext, r is a randomly chosen key, g is the generator, and n is the RSA modulus.

\begin{equation}
\label{eq:paillier_generic}
    {\varepsilon(m, r)} = ({ g^m \times r^n }) \quad mod \quad n^2
\end{equation}

The encryption of plaintext pairs $m_1$ and $m_2$ using random keys $r_1$ and $r_2$ is computed according to Equations \ref{eq:paillier_c1} and \ref{eq:paillier_c2}.

\begin{equation}
\label{eq:paillier_c1}
    {\varepsilon(m_1, r_1)} = ({ g^{m_1} \times {r_1}^n }) \quad mod \quad n^2
\end{equation}

\begin{equation}
\label{eq:paillier_c2}
    {\varepsilon(m_2, r_2)} = ({ g^{m_2} \times {r_2}^n }) \quad mod \quad n^2
\end{equation}

Subsequently, the multiplication of the ciphertexts is computed using Equation \ref{eq:paillier_c1_times_c2}.

\begin{equation}
\label{eq:paillier_c1_times_c2}
    {\varepsilon(m_1, r_1)} \times {\varepsilon(m_2, r_2)} =  ({ g^{m_1} \times {r_1}^n }) \times ({ g^{m_2} \times {r_2}^n }) \quad mod \quad n^2
\end{equation}

The multiplication can be rearranged as presented in Equations \ref{eq:paillier_c1_times_c2_reorganized} and \ref{eq:paillier_c1_times_c2_reorganized_2}.

\begin{equation}
\label{eq:paillier_c1_times_c2_reorganized}
    {\varepsilon(m_1, r_1)} \times {\varepsilon(m_2, r_2)} =  ({ g^{m_1} \times g^{m_2} \times {r_1}^n \times {r_2}^n }) \quad mod \quad n^2
\end{equation}

\begin{equation}
\label{eq:paillier_c1_times_c2_reorganized_2}
    {\varepsilon(m_1, r_1)} \times {\varepsilon(m_2, r_2)} =  ( g^{m_1 + m_2} \times {(r_1 \times r_2)}^n) \quad mod \quad n^2
\end{equation}

Alternatively, encrypting the sum of the plaintexts $m_1$ and $m_2$ with the random $r_1$ $\times$ $r_2$ will produce the same result, as demonstrated in Equation \ref{eq:paillier_encryption_of_m1_plus_m2}.

\begin{equation}
\label{eq:paillier_encryption_of_m1_plus_m2}
    {\varepsilon(m_1 + m_2, r_1 \times r_2)} =  ( g^{m_1 + m_2} \times {(r_1 \times r_2)}^n) \quad mod \quad n^2
\end{equation}

In conclusion, Paillier exhibits homomorphism with respect to addition, as demonstrated in Equation \ref{eq:paillier_summary}.

\begin{equation}
\label{eq:paillier_summary}
    {\varepsilon(m_1, r_1)} \times {\varepsilon(m_2, r_2)} = {\varepsilon(m_1 + m_2, r_1 \times r_2)}
\end{equation}

\subsubsubsection{Scalar Multiplication Feature:}

Although the Paillier cryptosystem is not homomorphic with respect to multiplication, it supports scalar multiplication, where a ciphertext can be multiplied by a known constant. Raising a ciphertext to the power of a constant yields the same result, as shown in Equations \ref{eq:paillier_scalar_1} and \ref{eq:paillier_scalar_2}.

\begin{equation}
\label{eq:paillier_scalar_1}
    {\varepsilon(m_1, r_1)}^k = ({ g^{m_1} \times {r_1}^n })^k \quad mod \quad n^2
\end{equation}

\begin{equation}
\label{eq:paillier_scalar_2}
    {\varepsilon(m_1, r_1)}^k = ({ g^{m_1 \times k} \times {r_1}^{n \times k} }) \quad mod \quad n^2
\end{equation}

On the other hand, multiplying $m_1$ by k with the random key $r_1$ raised to the k-th power will yield the same result, as shown in Equation \ref{eq:paillier_scalar_3}.

\begin{equation}
\label{eq:paillier_scalar_3}
    {\varepsilon(m_1 \times k, r_1^k)} = ({ g^{m_1 \times k} \times {r_1}^{n \times k} }) \quad mod \quad n^2
\end{equation}

After decryption, only the first argument of the encryption function will be recovered, which is the plaintext multiplied by the constant value.

\subsection{Damgård-Jurik}

The Damgård-Jurik cryptosystem is a generalized version of the Paillier algorithm, introduced by Ivan Damgård and Mads Jurik in 2001 \cite{damgard}. While Paillier performs computations modulo $n^2$, the Damgård-Jurik system operates modulo $n^{s+1}$, where n is an RSA modulus. In this sense, Paillier is a special case of Damgård-Jurik when s = 1. Like Paillier, the Damgård-Jurik cryptosystem relies on the difficulty of computing n-th residue classes. The generalized encryption calculation of the cryptosystem is shown in Equation \ref{eq:damgard_generic}.

\begin{equation}
\label{eq:damgard_generic}
    {\varepsilon(m, r)} = ({ g^m \times r^{n^s} }) \quad mod \quad n^{s+1}
\end{equation}

Then, the encryption of plaintext pairs $m_1$ and $m_2$ with random keys $r_1$ and $r_2$ will be calculated using Equations \ref{eq:damgard_c1} and \ref{eq:damgard_c2}.

\begin{equation}
\label{eq:damgard_c1}
    {\varepsilon(m_1, r_1)} = ({ g^{m_1} \times {r_1}^{n^s} }) \quad mod \quad n^{s+1}
\end{equation}

\begin{equation}
\label{eq:damgard_c2}
    {\varepsilon(m_2, r_2)} = ({ g^{m_2} \times {r_2}^{n^s} }) \quad mod \quad n^{s+1}
\end{equation}

Subsequently, the multiplication of ciphertexts will be computed using the equation.

\begin{equation}
\label{eq:damgard_c1_times_c2}
    {\varepsilon(m_1, r_1)} \times {\varepsilon(m_2, r_2)} =  ({ g^{m_1} \times {r_1}^{n^s} }) \times ({ g^{m_2} \times {r_2}^{n^s} }) \quad mod \quad n^{s+1}
\end{equation}

The multiplication can be rearranged as demonstrated in Equations \ref{eq:damgard_c1_times_c2_reorganized} and \ref{eq:damgard_c1_times_c2_reorganized_2}.

\begin{equation}
\label{eq:damgard_c1_times_c2_reorganized}
    {\varepsilon(m_1, r_1)} \times {\varepsilon(m_2, r_2)} =  ({ g^{m_1} \times g^{m_2} \times {r_1}^{n^s} \times {r_2}^{n^s} }) \quad mod \quad n^{s+1}
\end{equation}

\begin{equation}
\label{eq:damgard_c1_times_c2_reorganized_2}
    {\varepsilon(m_1, r_1)} \times {\varepsilon(m_2, r_2)} =  ( g^{m_1 + m_2} \times {(r_1 \times r_2)}^{n^s}) \quad mod \quad n^{s+1}
\end{equation}

On the other hand, encrypting the sum of plaintexts $m_1$ and $m_2$ with the random key $r_1$ $\times$ $r_2$ will yield the same result, as shown in Equation \ref{eq:damgard_encryption_of_m1_plus_m2}.

\begin{equation}
\label{eq:damgard_encryption_of_m1_plus_m2}
    {\varepsilon(m_1 + m_2, r_1 \times r_2)} =  ( g^{m_1 + m_2} \times {(r_1 \times r_2)}^{n^s}) \quad mod \quad n^{s+1}
\end{equation}

In conclusion, the Damgård-Jurik cryptosystem is homomorphic with respect to addition, as shown in Equation \ref{eq:damgard_summary}.

\begin{equation}
\label{eq:damgard_summary}
    {\varepsilon(m_1, r_1)} \times {\varepsilon(m_2, r_2)} = {\varepsilon(m_1 + m_2, r_1 \times r_2)}
\end{equation}

\subsubsubsection{Scalar Multiplication Feature:}

Although the Damgård-Jurik cryptosystem is not multiplicatively homomorphic, it supports scalar multiplication. The k-th power of a ciphertext, where k is a constant, can be calculated as shown in Equations \ref{eq:damgard_scalar_1} and \ref{eq:damgard_scalar_2}.

\begin{equation}
\label{eq:damgard_scalar_1}
    {\varepsilon(m_1, r_1)}^k = ({ g^{m_1} \times {r_1}^(n^s) })^k \quad mod \quad n^{s+1}
\end{equation}

\begin{equation}
\label{eq:damgard_scalar_2}
    {\varepsilon(m_1, r_1)}^k = ({ g^{m_1 \times k} \times {r_1}^{(n^s \times k)} }) \quad mod \quad n^{s+1}
\end{equation}

On the other hand, encrypting $m_1$ multiplied by k with the random key $r_1$ raised to the power of k will yield the same result, as shown in Equation \ref{eq:damgard_scalar_3}.

\begin{equation}
\label{eq:damgard_scalar_3}
    {\varepsilon(m_1 \times k, r_1^k)} = ({ g^{m_1 \times k} \times {r_1}^{(k \times n^s)} }) \quad mod \quad n^{s+1}
\end{equation}

The decryption process will restore only the first argument. The second argument, which contains the random number, will be completely discarded.

\subsection{Okamoto-Uchiyama}

The Okamoto-Uchiyama cryptosystem was invented by Tatsuaki Okamoto and Shigenori Uchiyama in 1998 \cite{okamoto}. The general encryption rule of the Okamoto-Uchiyama cryptosystem is shown in Equation \ref{eq:okamoto_generic}.

\begin{equation}
\label{eq:okamoto_generic}
    {\varepsilon(m, r)} = ({ g^m \times h^r }) \quad mod \quad n
\end{equation}

Then, the encryption of plaintext pairs $m_1$ and $m_2$ with random keys $r_1$ and $r_2$ will be calculated using Equations \ref{eq:okamoto_c1} and \ref{eq:okamoto_c2}.

\begin{equation}
\label{eq:okamoto_c1}
    {\varepsilon(m_1, r_1)} = ({ g^{m_1} \times {h}^{r_1} }) \quad mod \quad n
\end{equation}

\begin{equation}
\label{eq:okamoto_c2}
    {\varepsilon(m_2, r_2)} = ({ g^{m_2} \times {h}^{r_2} }) \quad mod \quad n
\end{equation}

Subsequently, the multiplication of ciphertexts will be computed using Equation \ref{eq:okamoto_c1_times_c2}.

\begin{equation}
\label{eq:okamoto_c1_times_c2}
    {\varepsilon(m_1, r_1)} \times {\varepsilon(m_2, r_2)} 
    =
    ({ g^{m_1} \times {h}^{r_1} }) \times ({ g^{m_2} \times {h}^{r_2} }) \quad mod \quad n
\end{equation}

The multiplication can be rearranged as demonstrated in Equations \ref{eq:okamoto_c1_times_c2_reorganized} and \ref{eq:okamoto_c1_times_c2_reorganized_2}.

\begin{equation}
\label{eq:okamoto_c1_times_c2_reorganized}
    {\varepsilon(m_1, r_1)} \times {\varepsilon(m_2, r_2)} 
    =
    ({ g^{m_1} \times g^{m_2} \times {h}^{r_1} \times {h}^{r_2} }) \quad mod \quad n
\end{equation}

\begin{equation}
\label{eq:okamoto_c1_times_c2_reorganized_2}
    {\varepsilon(m_1, r_1)} \times {\varepsilon(m_2, r_2)} 
    =
    ({ g^{m_1 + m_2} \times {h}^{r_1 + r_2} }) \quad mod \quad n
\end{equation}

On the other hand, encrypting the sum of plaintexts $m_1$ and $m_2$ with the random key $r_1$+$r_2$ will yield the same result, as shown in Equation \ref{eq:okamoto_m1_plus_m2_encrypted}.

\begin{equation}
\label{eq:okamoto_m1_plus_m2_encrypted}
    {\varepsilon(m_1 + m_2, r_1 + r_2)} 
    =
    ({ g^{m_1 + m_2} \times {h}^{r_1 + r_2} }) \quad mod \quad n
\end{equation}

In conclusion, the Okamoto-Uchiyama cryptosystem is homomorphic with respect to addition, as shown in Equation \ref{eq:okamoto_summary}.

\begin{equation}
\label{eq:okamoto_summary}
    {\varepsilon(m_1, r_1)} \times {\varepsilon(m_2, r_2)}
    =
    {\varepsilon(m_1 + m_2, r_1 + r_2)}
\end{equation}

\subsubsubsection{Scalar Multiplication Feature:}

The Okamoto-Uchiyama cryptosystem supports scalar multiplication, as shown in Equations \ref{eq:okamoto_scalar_1} and \ref{eq:okamoto_scalar_2}.

\begin{equation}
\label{eq:okamoto_scalar_1}
    {\varepsilon(m_1, r_1)}^k = ({ g^{m_1} \times {h}^{r_1} })^k \quad mod \quad n
\end{equation}

\begin{equation}
\label{eq:okamoto_scalar_2}
    {\varepsilon(m_1, r_1)}^k = ({ g^{m_1 \times k} \times {h}^{r_1 \times k} }) \quad mod \quad n
\end{equation}

On the other hand, encrypting $m_1$ multiplied by k with the random key $r_1$ raised to the power of k will yield the same result, as shown in Equation \ref{eq:damgard_scalar_3}.

\begin{equation}
\label{eq:okamoto_scalar_3}
    {\varepsilon(m_1 \times k, r_1 \times k)} = ({ g^{m_1 \times k} \times {h}^{r_1 \times k} }) \quad mod \quad n
\end{equation}

Once decrypted, the random number in the second argument is ignored.

\section{Vector Similarity}

Cosine similarity between two n-dimensional vectors is calculated by dividing the dot product of two n-dimensional vectors by the product of their magnitudes or lengths. It measures the cosine of the angle between those two vectors, which indicates how similar they are. It is basically shown in the Formula \ref{eq:cosine}.

\begin{equation}
\label{eq:cosine}
    {\theta(\mathbf{\alpha}, \mathbf{\beta}) = \frac{\mathbf{\alpha} \cdot \mathbf{\beta}}{\|\mathbf{\alpha}\| \|\mathbf{\beta}\|}} 
\end{equation}

To simplify the cosine similarity formula, if the source and target vectors are first normalized using the L2 Euclidean norm as shown in Equations \ref{eq:alpha_norm} and \ref{eq:beta_norm}, the cosine similarity formula reduces to the dot product of the two normalized vectors, as shown in Equation \ref{eq:cosine_updated}.

\begin{equation}
\label{eq:alpha_norm}
    { \mathbf{\hat{\alpha}} = \frac{\alpha}{\|\mathbf{\alpha}\|}} 
\end{equation}

\begin{equation}
\label{eq:beta_norm}
    { \mathbf{\hat{\beta}} = \frac{\beta}{\|\mathbf{\beta}\|}} 
\end{equation}

\begin{equation}
\label{eq:cosine_updated}
    {\theta(\mathbf{\alpha}, \mathbf{\beta}) = \mathbf{\hat{\alpha}} \cdot \mathbf{\hat{\beta}} } 
\end{equation}

We will explain how to implement this formula with PHE in the next session.

\section{Vector Similarity Computation with PHE}

The Two-Tower Architecture is a framework that utilizes two distinct branches, each responsible for encoding different inputs into vector embeddings. These embeddings are then projected into a shared space, where a similarity function (e.g., cosine similarity or Euclidean distance) measures their proximity. One tower encodes the query input (e.g., a recently captured image of a user’s face from a mobile device), while the other encodes candidate inputs (e.g., stored facial images of a target individual). This design facilitates efficient and scalable similarity searches, which are commonly employed in applications such as face verification, recommendation systems, and information retrieval.

\begin{figure}[H]
    \centering
    \includegraphics[width=0.95\textwidth]{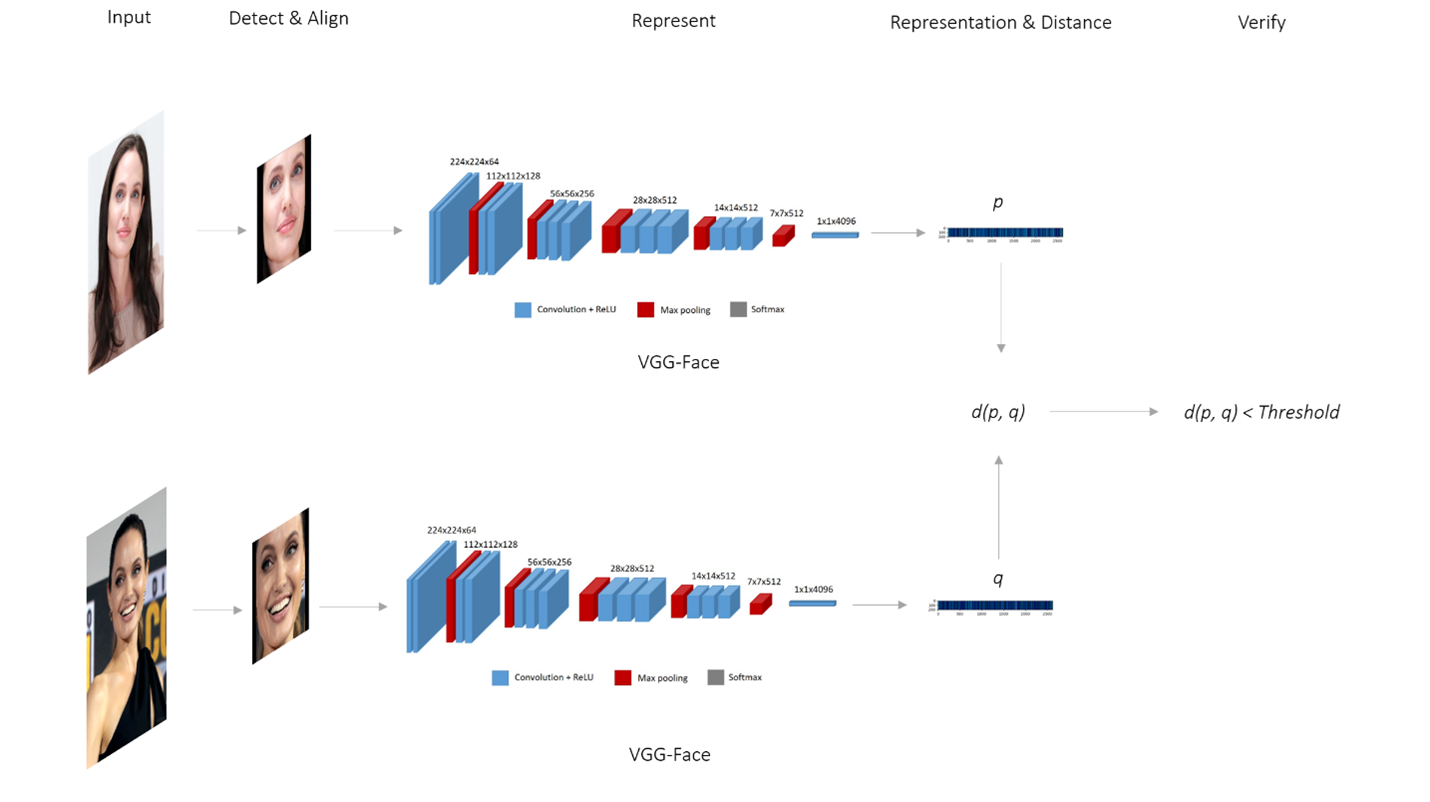}
    \caption{Facial Recognition Pipeline as a Two Tower Architecture}
    \label{fig:pipeline}
\end{figure}

A facial database containing millions of identities, where each identity is associated with a single facial image. On the server side, vector embeddings for all facial images must be precomputed, which is a one-time process. When a user attempts to log in, their face is captured by the mobile device, which generates a vector embedding using the same facial recognition model (e.g., FaceNet). The system then compares the newly generated vector embedding of the captured face with the pre-stored embeddings using a similarity function (e.g., cosine similarity or Euclidean distance). If the similarity score exceeds a predefined threshold, the system determines that the embeddings correspond to the same individual.

In this setup, the server-side operation occurs within the user tower branch, where vector embeddings for each identity are precomputed in advance, while the mobile device operates within the item tower branch, handling real-time input. A key component of this framework involves encrypting the vector embeddings generated on the server side.

Cosine similarity is calculated by multiplying the corresponding components of two normalized vectors and summing the results. A critical observation is that the vector embedding from the server is already encrypted, while the vector embedding from the mobile device remains in plaintext. Additively homomorphic encryption algorithms, which support the addition of encrypted values and the multiplication of encrypted values by scalars, enable this computation.

Specifically, the corresponding dimension values of the encrypted vector from the server can be multiplied element-wise with the plaintext vector from the mobile device. The result is an encrypted vector of values for each dimension. Subsequently, the encrypted dimensions of this new vector can be summed—enabled by the additively homomorphic encryption—yielding the encrypted cosine similarity. This method enables the secure computation of cosine similarity while preserving data privacy. Algorithm \ref{alg:compute_similarity} explains how to compute encrypted cosine similarity with an encrypted vector embedding and plain vector embedding pair.

\begin{algorithm}[H]
\caption{Encrypted Vector Similarity Computation}
\label{alg:compute_similarity}
\begin{algorithmic}[1]
\Require Source vector's encrypted representation $encrypted\_source\_embedding$, target vector's plain representation $target\_embedding$, both representations have $n$ dimensions
\Ensure Encrypted cosine similarity stored in $encrypted\_similarity$.

\State $encrypted\_similarity = encrypted\_source\_embedding[0] \times target\_embedding[0]$
\For{$i \gets 1$ \textbf{to} $n$}
    \State $encrypted\_similarity  += encrypted\_source\_embedding[i] \times target\_embedding[i]$
\EndFor
\State \Return $encrypted\_similarity$
\end{algorithmic}
\end{algorithm}

The code snippet \ref{encrypted_vector_similarity} shows the implementation of encrypted vector similarity computation in Python with LightPHE.

\begin{minipage}{\linewidth}
\begin{lstlisting}[frame=tb, caption=Encrypted Vector Similarity Computation with LightPHE, label=encrypted_vector_similarity, language=Python]
from lightphe import LightPHE
from lightphe.models.Tensor import EncryptedTensor

def onprem(source_embedding: List[float]) -> EncryptedTensor:
    # build an additively homomorphic cryptosystem
    cs = LightPHE(algorithm_name="Paillier", precision=19)

    # export keys
    cs.export_keys("secret.txt")
    cs.export_keys("public.txt", public=True)

    # encrypt the embedding
    encrypted_source_embedding = cs.encrypt(source_vector)

    return encrypted_source_embedding

def cloud(
    encrypted_source_embedding: EncryptedTensor,
    target_embedding: List[float]
) -> EncryptedTensor:
    # restore the cryptosystem with public key
    cs = LightPHE(algorithm_name="Paillier", precision=19 key_file="public.txt")

    # find dot product of encrypted embedding and plain embedding
    encrypted_cosine_similarity = encrypted_source_embedding @ target_embedding

    # confirm that cloud cannot decrypt it even though it is calculated by cloud
    with pytest.raises(ValueError, match="must have private key"):
        cs.decrypt(encrypted_cosine_similarity)

    return encrypted_cosine_similarity

def proof_of_work(
    encrypted_cosine_similarity: EncryptedTensor,
    source_embedding: List[float],
    target_embedding: List[float]
):
    # restore the built cryptosystem on-prem with secret key
    cs = LightPHE(algorithm_name="Paillier", precision=19, key_file="secret.txt")

    # restore cosine similarity
    cosine_similarity = cs.decrypt(encrypted_cosine_similarity)[0]
    
    expected_similarity = sum(
        x * y for x, y in zip(source_embedding, target_embedding)
    )

    return abs(cosine_similarity - expected_similarity) < 0.01

# define a plain vector embeddings (L2 normalized and all positive)
source_embedding = [7.11, 5.22, 5.33, 2.44, 3.55, 4.66]
target_embedding = [5.66, 3.77, 2.88, 4, 0, 5.99]

encrypted_source_embedding = onprem(source_embedding)
encrypted_cosine_similarity = cloud(encrypted_source_embedding, target_embedding)
assert proof_of_work(
    encrypted_cosine_similarity, source_embedding, target_embedding
) is True
\end{lstlisting}
\end{minipage}
\newline
\newline

\section{Experiments}

\begin{sidewaystable}[]
\begin{tabular}{llll|llllllll}
\hline
\multicolumn{1}{c}{cryptosystem} & \multicolumn{1}{c}{security} & \multicolumn{1}{c}{model} & \multicolumn{1}{c}{ndim} & \multicolumn{1}{c}{\begin{tabular}[c]{@{}c@{}}encrypt\\ (it/s)\end{tabular}} & \multicolumn{1}{c}{\begin{tabular}[c]{@{}c@{}}homomorphic\\ (it/s)\end{tabular}} & \multicolumn{1}{c}{\begin{tabular}[c]{@{}c@{}}decrypt\\ (it/s)\end{tabular}} & \multicolumn{1}{c}{loss} & \multicolumn{1}{c}{\begin{tabular}[c]{@{}c@{}}secret\\ (MB)\end{tabular}} & \multicolumn{1}{c}{\begin{tabular}[c]{@{}c@{}}public\\ (MB)\end{tabular}} & \multicolumn{1}{c}{\begin{tabular}[c]{@{}c@{}}embedding\\ (MB)\end{tabular}} & \multicolumn{1}{c}{\begin{tabular}[c]{@{}c@{}}similarity\\ (MB)\end{tabular}} \\
\hline
Paillier & 80 & FaceNet & 128& 1.6569 & 8.0720 & 42.336& 9.36E-14 & 0.0009& 0.0006& 0.1308 & 0.0012\\
Damgård-Jurik& 80 & FaceNet & 128& 1.0646 & 3.9611 & 19.188& 9.36E-14 & 0.0009& 0.0006& 0.1935 & 0.0017\\
Okamoto-Uchiyama & 80 & FaceNet & 128& 1.8827 & 14.621& 146.58 & 9.36E-14 & 0.0017& 0.0013& 0.0975 & 0.0009\\
Paillier & 80 & FaceNet& 512& 0.6506 & 2.2145 & 40.957& 9.68E-14 & 0.0009& 0.0006& 0.5227 & 0.0012\\
Damgård-Jurik& 80 & FaceNet& 512& 0.2888 & 0.9741 & 15.183& 9.68E-14 & 0.0009& 0.0006& 0.7736 & 0.0017\\
Okamoto-Uchiyama & 80 & FaceNet& 512& 0.5529 & 3.0917 & 118.94 & 9.68E-14 & 0.0017& 0.0013& 0.3895 & 0.0009\\
Paillier & 80 & VGG-Face& 4096 & 0.4774 & 2.2074 & 41.497& 1.15E-07 & 0.0009& 0.0006& 4.1805 & 0.0012\\
Damgård-Jurik& 80 & VGG-Face& 4096 & 0.2416 & 1.0426 & 18.486& 1.15E-07 & 0.0009& 0.0006& 6.1873 & 0.0017\\
Okamoto-Uchiyama & 80 & VGG-Face& 4096 & 0.3978 & 3.2195 & 145.37 & 1.15E-07 & 0.0017& 0.0013& 3.1151 & 0.0009\\
Paillier & 112& FaceNet & 128& 0.3979 & 2.4139 & 5.8880 & 9.36E-14 & 0.0018& 0.0012& 0.2559 & 0.0022\\
Damgård-Jurik& 112& FaceNet & 128& 0.2095 & 1.2027 & 2.8951 & 9.36E-14 & 0.0018& 0.0012& 0.3802 & 0.0031\\
Okamoto-Uchiyama & 112& FaceNet & 128& 0.4504 & 4.2569 & 20.584& 9.36E-14 & 0.0033& 0.0027& 0.1934 & 0.0017\\
Paillier & 112& FaceNet& 512& 0.1101 & 0.6192 & 5.7851 & 9.68E-14 & 0.0018& 0.0012& 1.0231 & 0.0022\\
Damgård-Jurik& 112& FaceNet& 512& 0.0558 & 0.3055 & 2.8934 & 9.68E-14 & 0.0018& 0.0012& 1.5205 & 0.0031\\
Okamoto-Uchiyama & 112& FaceNet& 512& 0.1233 & 1.0555 & 20.246& 9.68E-14 & 0.0033& 0.0027& 0.7730 & 0.0017\\
Paillier & 112& VGG-Face& 4096 & 0.1143 & 0.6555 & 5.7180 & 1.15E-07 & 0.0018& 0.0012& 8.1835 & 0.0022\\
Damgård-Jurik& 112& VGG-Face& 4096 & 0.0621 & 0.3252 & 2.8947 & 1.15E-07 & 0.0018& 0.0012& 12.163& 0.0031\\
Okamoto-Uchiyama & 112& VGG-Face& 4096 & 0.1281 & 1.1339 & 20.800& 1.15E-07 & 0.0033& 0.0027& 6.1830 & 0.0017\\
\hline
TenSEAL (p=$2^{13}$, q=200) & 128& FaceNet & 128& 143.90 & 34.374& 432.78 & 0.001548 & 0.8964& 45.106 & 0.4251 & 0.2993\\
TenSEAL (p=$2^{13}$, q=200) & 128& FaceNet& 512& 140.06 & 26.642& 392.87 & 0.000258& 0.8964& 45.106 & 0.4251 & 0.2993\\
TenSEAL (p=$2^{13}$, q=200) & 128& VGG-Face& 4096 & 127.14 & 24.335& 434.94 & 0.001025& 0.8964& 45.106 & 0.4251 & 0.2993\\
TenSEAL (p=$2^{14}$, q=422) & 128& FaceNet & 128& 43.708& 4.5354 & 69.160& 1.344621& 3.5502& 451.01& 2.1820 & 1.8487\\
TenSEAL (p=$2^{14}$, q=422)& 128& FaceNet& 512& 38.927& 3.3899 & 61.121& 0.951100& 3.5502& 451.01& 2.1820 & 1.8487\\
TenSEAL (p=$2^{14}$, q=422)& 128& VGG-Face& 4096 & 33.941& 2.8523 & 66.155 & 3.157271& 3.5502& 451.01& 2.1820 & 1.8487 \\
\hline
\end{tabular}
\label{tab:experiments}
\end{sidewaystable}

The experimental results in Table \ref{tab:experiments} demonstrate the varying strengths and trade-offs between FHE (TenSEAL) and PHE (Paillier, Damgård-Jurik, and Okamoto-Uchiyama) for encrypted vector similarity search using face recognition models (FaceNet and VGG-Face) across different input dimensions (128, 512, 4096).

\textbf{Encryption, Decryption, and Homomorphic Operations:} TenSEAL and FHE clearly outperform all traditional cryptosystems in encryption, decryption, and homomorphic operations, making it the most efficient cryptosystem for privacy-preserving computations in face recognition tasks. TenSEAL achieves encryption speeds that are significantly faster compared to Paillier, Damgård-Jurik, and Okamoto-Uchiyama. For example, for the FaceNet model with a dimension of 128, TenSEAL performs encryption 143.91 items per second, which is several times faster than the best-performing traditional cryptosystem (Okamoto-Uchiyama), which achieves only 1.88 encryptions per second. The homomorphic operation performance follows a similar trend, with TenSEAL achieving up to 34.37 operations per second, which is 2.4 times faster than the best traditional cryptosystem in this category. The decryption performance is also favorable for TenSEAL, albeit slightly slower than Okamoto-Uchiyama. Despite this, TenSEAL still remains competitive and faster than other cryptosystems in most configurations.

\textbf{Secret and Public Key Sizes, Embedding Size, and Similarity Size:} On the other hand, Paillier, Damgård-Jurik, and Okamoto-Uchiyama are more efficient when it comes to the secret and public key sizes, encrypted embedding size, and encrypted similarity size. These traditional cryptosystems have smaller memory footprints compared to TenSEAL, making them more suitable for resource-constrained environments such as IoT and edge devices or mobile phones. For instance, the secret and public keys for TenSEAL are significantly larger than those for Paillier and Damgård-Jurik. Paillier, for example, requires only 0.0009 MB for the secret and 0.0006 MB for public keys, while TenSEAL requires up to 3.55 MB for secret key file and 451 MB for public key file. Furthermore, TenSEAL’s embedding and similarity sizes are also considerably larger, particularly for configurations with larger parameters. These increased memory requirements should be taken into account when selecting a cryptosystem for deployment on devices with limited memory.

\textbf{Loss Comparison:} A critical point of comparison is the loss, which refers to the difference between the original similarity value computed from plain embeddings and the restored similarity value obtained after computing the dot product of an encrypted-plain embedding pair, followed by decryption. Paillier, Damgård-Jurik, and Okamoto-Uchiyama consistently maintain low loss values, ensuring high fidelity in the encrypted computations. The loss for these systems remains extremely small, often in the range of $10^-14$ or lower. In contrast, TenSEAL exhibits an increasing loss, which grows as the parameter size increases. For instance, in the configuration with p=$2^{14}$ and q=400, the loss for TenSEAL reaches 3.15, which is much higher compared to the traditional systems. This increase in loss may have significant implications, particularly in applications like classification (e.g. face recognition), where precision is critical. The increasing loss could potentially lead to misclassification after decryption, affecting the overall performance of the system.

\textbf{Encrypted Embedding and Encrypted Similarity Sizes:} In terms of encrypted embedding size and encrytped similarity size, the traditional cryptosystems (Paillier, Damgård-Jurik, and Okamoto-Uchiyama) perform well with small memory requirements. However, TenSEAL requires higher memory (e.g., 451 MB for public key file size).

\textbf{Summary of Performance Differences:}

\textbf{Encryption Speed:} TenSEAL outperforms traditional cryptosystems by a factor of 80 to 100 times, achieving speeds up to 143.91 encryptions per second for FaceNet with a dimension of 128.

\textbf{Homomorphic Operation Speed:} TenSEAL is 2-3 times faster than the best-performing traditional cryptosystem (Okamoto-Uchiyama), achieving up to 34.37 homomorphic operations per second.

\textbf{Decryption Speed:} Okamoto-Uchiyama offers faster decryption speeds (up to 146.58 per second) compared to TenSEAL, but TenSEAL’s decryption speeds are still competitive.

\textbf{Key Sizes and Memory Usage:} Paillier, Damgård-Jurik, and Okamoto-Uchiyama have significantly smaller secret/public key sizes, encrypted embedding sizes and encrypted similarity size, with memory usage up to 10 times smaller than TenSEAL.

\textbf{Loss:} TenSEAL's loss increases with the size of the parameters, which could lead to misclassifications, whereas the traditional cryptosystems maintain minimal loss.

In conclusion, TenSEAL provides superior performance in encryption, decryption, and homomorphic operations, making it an excellent choice for scenarios where computational efficiency is paramount. However, the increasing loss with TenSEAL may be a concern for high-precision applications like face recognition, where small discrepancies in the embeddings could lead to misclassifications. For environments with limited memory or where embedding size and low loss are more critical, traditional cryptosystems like Paillier, Damgård-Jurik, and Okamoto-Uchiyama offer a better balance between performance and resource usage. The decision between these cryptosystems should be guided by the specific needs of the application, balancing the trade-offs between speed, memory usage, and the acceptable level of loss in encrypted computations.

\section{Conclusion}

This paper demonstrates the viability of using Partially Homomorphic Encryption (PHE) for privacy-preserving encrypted vector similarity search, with a specific focus on facial recognition. By leveraging PHE, we propose an efficient and scalable solution for securely computing cosine similarity between encrypted vector embeddings, addressing the significant computational overhead associated with Fully Homomorphic Encryption (FHE). Our approach introduces a normalization-based transformation that enables the use of dot products as a proxy for cosine similarity, overcoming the inherent limitations of PHE.

Through extensive experiments on the Labeled Faces in the Wild (LFW) dataset, we show that PHE can effectively perform encrypted vector similarity search while being computationally more efficient, faster, and requiring less memory than FHE. The results indicate that PHE-based methods significantly outperform FHE in practical applications, particularly in resource-constrained environments such as edge devices and mobile platforms. While FHE offers greater flexibility for fully arbitrary computations, the trade-offs in efficiency, computational overhead, and key management make PHE a more practical choice for many real-world privacy-preserving applications.

Our findings highlight the potential of PHE as a lightweight yet secure alternative to FHE, enabling encrypted similarity search in domains ranging from facial recognition to recommendation systems and large language models. Future work can explore the integration of more advanced encryption schemes, further optimization of the PHE-based methods, and the extension of our approach to other similarity measures. Ultimately, this study paves the way for more efficient and scalable privacy-preserving machine learning applications in various sensitive domains.

\newpage

\bibliographystyle{unsrt}

\end{document}